\documentclass{JAC2003}

\usepackage{graphicx}

\begin{document}
\title{PHYSICS POTENTIAL OF SPS UPGRADE 
\\  IN REGARD TO BETA/EC BEAMS}

\author{Jos\'e Bernab\'eu, ~Catalina Espinoza,  IFIC, Univ.~Valencia-CSIC}

\maketitle

\begin{abstract}
\noindent The goal for future neutrino facilities is the determination  of the $[U_{e3}]$ mixing and CP violation in neutrino oscillations. This will require precision experiments with a very intense 
neutrino source. With this objective the creation of neutrino beams from the radioactive decay of boosted ions by the SPS of CERN from either beta or electron capture transitions has been propossed. We discuss the capabilities of such facilities as a function of the energy of the boost and the baseline for the detector. We conclude that the SPS upgrade to 1000 GeV is crucial to have a better sensitivity to CP violation if it is accompanied by a longer baseline. We compare the physics potential for two different configurations. In the case of beta beams,  with the same boost for both $\beta^+$ (neutrinos) and $\beta^-$ (antineutrinos), the two setups are: I) $\gamma=120$, $L=130$~Km (Frejus); II) $\gamma=330$, $L=650$~Km (Canfranc). In the case of monochromatic EC beams we exploit the energy dependence of neutrino oscillations to separate out the two parameters  $U(e3)$ and the CP phase $\delta$. Setup I runs at $\gamma=90$ and $\gamma=195$ (maximum achievable at present SPS) to Frejus, whereas Setup II runs at  $\gamma=195$ and $\gamma=440$ (maximum achievable at upgraded SPS) to Canfranc. The main conclusion is that, whereas the gain in the determination of $U(e3)$ is rather modest, setup II provides  much better sensitivity to CP violation.
\end{abstract}

\section{1.~Introduction}
\noindent Neutrinos are very elusive particles that are difficult to detect. Even so, physicists have
over the last decades successfully studied neutrinos from a wide variety of sources, either natural,
such as the sun and cosmic objects, or manmade, such as nuclear power plants or accelerated beams.
 Spectacular results have been obtained in the last few years for the flavour mixing of neutrinos
 obtained from atmospheric, solar, reactor and accelerator sources and interpreted in terms of the
 survival probabilities for the beautiful quantum phenomenon of neutrino oscillations
\cite{fukuda,ahmad}. The weak interaction eigenstates  $\nu_{\alpha}$ $(\alpha = e,\mu, \tau)$
 are written in terms of mass eigenstates
$\nu_k$ $(k=1,2,3)$ as  $\nu_{\alpha} = \sum_k U_{\alpha k} (\theta_{12}, \theta_{23},
\theta_{13};\delta ) \nu_k$, where $\theta_{ij}$ are the mixing angles among the three neutrino
families and $\delta$  is the $CP$ violating phase. Neutrino mass differences and the mixings for
the atmospheric $\theta_{23}$ and solar $\theta_{12}$ sectors have thus been determined. The third connecting mixing $\vert U_{e3} \vert$ is bounded as  $\theta_{13}\le 10^{\circ}$   from the CHOOZ
reactor experiment \cite{apollonio}. 
The third angle, $\theta_{13}$,   as well as the 
CP-violating phases $\delta$, remain  thus undetermined. Besides the approved experiments Double CHOOZ \cite{Lasserre:2004vt}, T2K \cite{t2k} and NOVA \cite{Ayres:2004js},  a number
of experimental facilities to significantly improve on present
sensitivity have been discussed in the literature:
neutrino factories (neutrino beams from boosted-muon 
decays)~\cite{geer,drgh,nufact}, 
superbeams (very intense conventional neutrino beams)~\cite{JHF,NUMI,splcern,superbeam}
improved reactor experiments~\cite{reactor_deg} 
and more recently $\beta$-beams \cite{zucchelli}.
The original standard scenario for beta beams with lower $\gamma=60/100$ and  short baseline $L=130$~Km from CERN to Frejus with $^6He$ and $^{18}Ne$ ions could be improved using an electron capture facility for monochromatic neutrino beams \cite{Bernabeu:2005jh}. New proposals also include the high $Q$ value $^8Li$ and $^8Be$ isotopes in a $\gamma=100$ facility \cite{Rubbia:2006pi}. In this paper we discuss the physics reach that a high energy facility for both beta beams \cite{Burguet-Castell:2005pa} and EC beams may provide with the expected SPS upgrade at CERN.
In Section~2 we discuss the virtues of the suppressed oscillation channel $(\nu_e \to \nu_\mu)$ in order to have access to the parameters $\theta_{13}$ and $\delta$. The interest of energy dependence, as obtainable in the EC facility, is emphasized. In Section~3 we compare the beta beam capabilities at different energies and baselines using two ions, one for neutrinos, the other for antineutrinos. In Section~4 we present new results on the comparison between (low energies, short baseline) and (high energies, long baseline) configurations for an EC facility with a single ion. Section~5 gives our conclusions and outlook.

\section{2.~Suppressed Neutrino Oscillation}
\noindent The observation of $CP$ violation needs an experiment in which the emergence of another neutrino flavour
 is detected rather than the deficiency of the original flavour of the neutrinos. At the same time, the interference needed to generate CP-violating observables can be enhanced if both the atmospheric and solar components have a similar magnitude. This happens in the suppressed $\nu_e \to \nu_{\mu}$ transition. The appearance
 probability $P(\nu_e \to \nu_{\mu})$ as a function of the distance between source and detector $(L)$ is given by \cite{cervera}
\begin{eqnarray}\label{prob}
P({\nu_e \rightarrow \nu_\mu})  \simeq ~ 
s_{23}^2 \, \sin^2 2 \theta_{13} \, \sin^2 \left ( \frac{\Delta m^2_{13} \, L}{4E} \right ) \nonumber \\
  + ~   c_{23}^2 \, \sin^2 2 \theta_{12} \, \sin^2 \left( \frac{ \Delta m^2_{12} \, L}{4E} \right ) 
\nonumber \\
 + ~ \tilde J \, \cos \left ( \delta - \frac{ \Delta m^2_{13} \, L}{4E} \right ) \;
\frac{ \Delta m^2_{12} \, L}{4E} \sin \left ( \frac{  \Delta m^2_{13} \, L}{4E} \right ) \,,
\end{eqnarray}
where $\tilde J \equiv c_{13} \, \sin 2 \theta_{12} \sin 2 \theta_{23} \sin 2 \theta_{13}$.
   The three terms of Eq.~(\ref{prob}) correspond, respectively, to contributions from the
atmospheric and solar sectors and their interference. As seen, the $CP$ violating contribution
has to include all mixings and neutrino mass differences to become observable. The four measured parameters $(\Delta m_{12}^2,\theta_{12})$  and  $(\Delta m_{23}^2,\theta_{23})$ have been fixed throughout this paper to their mean values \cite{Gonzalez-Garcia:2004jd}. 

\begin{figure}[ht!]
\centering
\includegraphics[width=55mm, angle=-90]{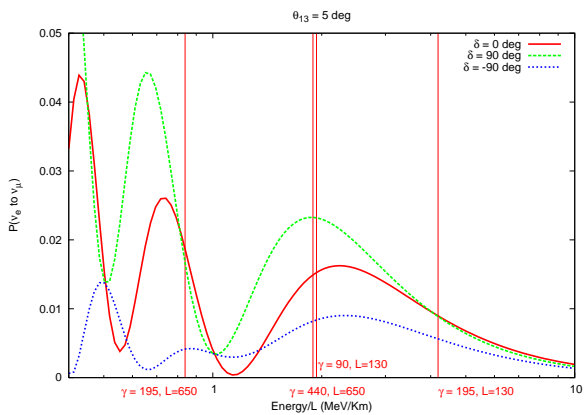}
\caption{The appearance probability $P(\nu_e \rightarrow \nu_{\mu})$ for neutrino oscillations
 as a function of the LAB energy E/L, with fixed  connecting
mixing. The three curves refer to different values of the CP violating phase $\delta$. The 
vertical lines are the energies of our simulation study in the EC facility.}
\label{proba}
\end{figure}

Neutrino oscillation phenomena are energy dependent (see Fig.\ref{proba}) for a fixed distance between source and detector, and the observation of this energy dependence would disentangle the two important parameters: whereas $\vert U_{e3} \vert$ gives the strength of the appearance probability, the $CP$ phase acts as a phase-shift in the interference pattern. These properties suggest the consideration of a facility able to study the detailed energy dependence by means of fine tuning of a boosted monochromatic neutrino beam. As shown below, in an electron capture facility the neutrino energy is dictated by the chosen boost
of the ion source and the neutrino beam luminosity is concentrated at a single
known energy which may be chosen at
will for the values in which the sensitivity for the $(\theta_{13}, \delta)$ parameters is higher. This is in contrast to beams with a continuous spectrum, where the intensity is shared between sensitive
and non sensitive regions. Furthermore, the definite energy would help in the control of both the systematics
and the detector background. In the beams with a continuous spectrum, the neutrino energy has to be reconstructed in the detector. In water-Cerenkov detectors, this reconstruction is made from supposed quasielastic events by measuring both the energy and direction of the charged lepton. This procedure suffers from non-quasielastic background, from kinematic deviations due to the nuclear Fermi momentum and from dynamical suppression due to exclusion effects \cite{Bernabeu72}.

From general arguments of CPT invariance and absence of absorptive parts the CP-odd probability is odd in time and then odd in the baseline L (formally). Vacuum oscillations are only a function of $E/L$ so that, at fixed $L$, the CP-odd probability is odd in the energy (formally). This proves that the study of neutrino oscillations in terms of neutrino energy will be able to separate out the CP phase $\delta$ from the mixing parameters. A control of this energy may be obtained from the choice of the boost in the EC facility with a single ion. In order for this concept to become operational it is necessary to combine it with the recent discovery of nuclei  far from the stability line, having super
 allowed spin-isospin transitions to a giant Gamow-Teller resonance kinematically accessible
 \cite{algora}. Thus the rare-earth nuclei above $^{146}Gd$ have a small enough half-life for
 electron capture processes. This is in contrast with the proposal of EC beams  with fully stripped long-lived ions \cite{Sato:2005ma}.  We discuss the option of short-lived ions \cite{Bernabeu:2005jh}.

\section{3.~Beta-Beam capabilities at different energies and baselines}
\noindent A first question to be answered is: Is the sensitivity to CP violation and $\theta_{13}$ changing with energy at fixed baseline? Fixing the baseline to CERN-Frejus ($L=130$~Km) \cite{Burguet-Castell:2005pa}, one notices that, for $\gamma>80$, the sensitivity to both $\theta_{13}$ and $\delta$ changes rather slowly because the flux at low energies in the continuous spectrum does not reduce significantly. Then it is not advantageous to increase the neutrino energy unless the baseline is  correspondingly scaled to remain close to the atmospheric oscillation maximum as suggested by the $E/L$ dependence.

With the present SPS of CERN the maximum energy reachable for the $^6He$ ion corresponds to $\gamma=150$. Fixing this value of $\gamma$ for both $^6He$ and  $^{18}Ne$ we may ask: Is the sensitivity to $\theta_{13}$ and $\delta$ changing with the baseline? Particularly for the CP phase $\delta$, $L=300$~Km is clearly favoured \cite{Burguet-Castell:2005pa}. There is neither an existing nor an envisaged laboratory at this particular distance from CERN.

\begin{figure}[htb]
\centering
\includegraphics*[width=65mm]{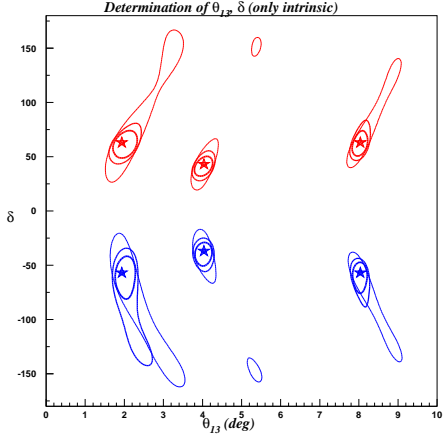}
\caption{Determination of $(\theta_{13}, \delta)$ for the Setups I) and II), as explained in the text.}
\label{fit-betabeams}
\end{figure}

Equipped with these previous results, it is of interest to make a comparison between the physics reach for two different Beta Beam Setups. With the same $\gamma$ for both neutrino and antineutrino sources, Setup I corresponds to $\gamma=120$, $L=130$~Km (Frejus), whereas Setup II is for $\gamma=330$, $L=650$~Km (Canfranc). Setup II needs the upgrade of the SPS until proton energies of $1000$~GeV. The associated determinations of $\theta_{13}$ and $\delta$ are presented in Fig.~\ref{fit-betabeams}. The main conclusion is that Setup II is clearly better for the CP violating phase. Not only the high energy Setup provides a better precision, but it is able to resolve the degeneracies. In \cite{Burguet-Castell:2005pa} one may find the associated sensitivities of these Setups for each parameter $\theta_{13}$ and $\delta$. As a consequence, a $R \& D$ effort to design Beta Beams for the upgraded  CERN SPS ($E_p=1000$~GeV) appears justified.

\section{4.~EC-Beam capabilities at different energies and baselines}
\begin{figure}[htb]
\centering
\includegraphics*[width=65mm]{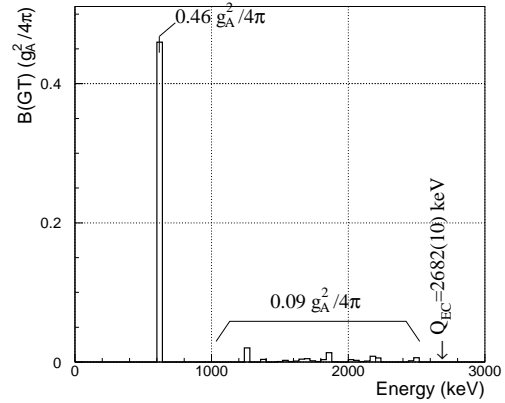}
\caption{Gamow-Teller strength distribution in the $EC/\beta^+$ decay of $^{148}Dy$.}
\label{dy}
\end{figure}

\noindent Electron Capture is the process in which an atomic electron is captured by a proton
of the nucleus leading to a nuclear state of the same mass number $A$, replacing the 
proton by a neutron, and a neutrino. Its probability amplitude is proportional to the atomic 
wavefunction at the origin, so that it becomes competitive with the nuclear $\beta^+$ decay 
at high $Z$. Kinematically, it is a two body decay of the atomic ion into a nucleus and the 
neutrino, so that the neutrino energy is well defined and given by the difference between
the initial and final atomic masses  $(Q_{EC})$ minus the excitation energy of the 
final nuclear state. In general, the high proton number $Z$ nuclear beta-plus decay $(\beta^+)$
 and electron-capture $(EC)$ transitions are very "forbidden", i.e., disfavoured, because the
energetic window open $Q_{\beta}/Q_{EC}$ does not contain the important Gamow-Teller strength
excitation seen in (p,n) reactions. There are a few cases, however, where the Gamow-Teller 
resonance can be populated (see Fig.\ref{dy}) having the occasion of a direct study of the "missing"
 strength. For the rare-earth nuclei above $^{146}Gd$, the filling of the intruder level $h_{11/2}$ 
for protons opens the possibility of a spin-isospin transition to the allowed level $h_{9/2}$
for neutrons, leading to a fast decay. Our studies for neutrino beam preparation have used the $^{150}Dy$ ion with half live of 7.2 min, a Branching Ratio to neutrino channels of $64\%$ (fully by EC) and neutrino energy of $1.4$~MeV in the C.M. frame as obtained from its decay to the single giant Gamow-Teller resonance in the daugther $^{150}Tb^*$. 
 A neutrino (of energy $E_0$) that emerges from radioactive decay in an accelerator will
 be boosted in energy. At the experiment, the measured energy distribution as a function
 of angle $(\theta)$ and Lorentz gamma $(\gamma)$ of the ion at the moment of decay can 
be expressed as $E = E_0 / [\gamma(1- \beta \cos{\theta})]$. The angle $\theta$ in the 
formula expresses the deviation between the actual neutrino detection and the ideal detector
 position in the prolongation of one of the long straight sections of the Decay Ring.
 The neutrinos are concentrated inside a narrow cone around the forward direction. If the ions
 are kept in the decay ring longer than the half-life, the energy distribution of the Neutrino 
Flux arriving to the detector in absence of neutrino oscillations is given by the Master Formula

\begin{eqnarray}\label{master}
\frac{d^2N_\nu}{ dS dE}
& = & \frac{1}{\Gamma} \frac{d^2\Gamma_\nu}{dS dE} N_{ions} \nonumber \\
& \simeq & \frac{\Gamma_\nu}{\Gamma} \frac{ N_{ions}}{\pi L^2} \gamma^2
\delta{\left(E - 2 \gamma E_0 \right)},
\end{eqnarray}
with a dilation factor $\gamma >> 1$. It is remarkable that the result is given only in terms of the branching ratio 
and the neutrino energy and independent of nuclear models. In Eq.~(\ref{master}), $N_{ions}$ is the total number of 
ions decaying to neutrinos. For an optimum choice with $E \sim L$ around the
 first oscillation maximum, Eq.~(\ref{master}) says that lower neutrino energies $E_0$ in the
 proper frame give higher neutrino fluxes.  The number of events will increase with higher
 neutrino energies as the cross section increases with energy. To conclude, in the forward 
direction the neutrino energy is fixed by the boost $E = 2 \gamma E_0$, with the entire neutrino 
flux concentrated at this energy. As a result, such a facility will measure the neutrino 
oscillation parameters by changing the $\gamma$'s of the decay ring (energy dependent measurement)
 and there is no need of energy reconstruction in the detector.

For the study of the physics reach associated with such a facility, we combine two different energies for the same $^{150}Dy$ ion using two Setups. In all cases we consider $10^{18}$ decaying ions/year, a water Cerenkov Detector with fiducial mass of $440$~Kton and both appearance ($\nu_{\mu}$) and disappearance ($\nu_e$) events. Setup I  is associated with a five year run at $\gamma=90$ (close to the minimum energy to avoid atmospheric neutrino background) plus a five year run at $\gamma=195$ (the maximum energy achievable at present SPS), with a baseline $L=130$~Km from CERN to Frejus. The results for Setup I are going to be compared with those for Setup II, associated with a five year run at $\gamma=195$ plus a five year run at $\gamma=440$ (the maximum achievable at the upgraded SPS with Proton energy of $1000$~GeV), with a baseline $L=650$~Km from CERN to Canfranc.

\begin{figure}[htb]
\centering
\includegraphics*[width=75mm]{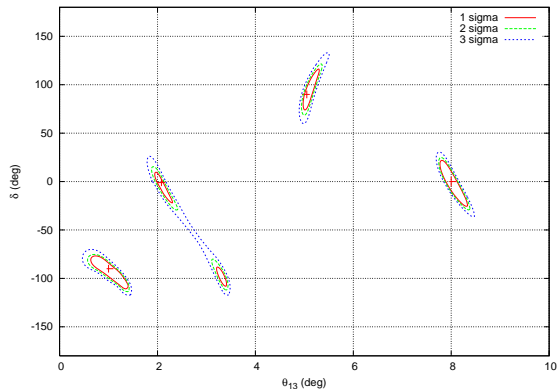}
\caption{Setup I. Fit for $(\theta_{13}, \delta)$ from statistical distribution.}
\label{fit-setupI}
\end{figure}

For the Setup I we generate the statistical distribution of events from  assumed values of $\theta_{13}$ and $\delta$. The corresponding fit is shown in Fig.~\ref{fit-setupI} for selected values of $\theta_{13}$ from $6^o$ to $1^o$ and covering a few values of the CP phase $\delta$. As observed, the principle of an energy dependent measurement (illustrated here with two energies) is working to separate out the two parameters. With this configuration the precision obtainable for the mixing (even at 1 degree) is much better than that for the CP phase.

\begin{figure}[htb]
\centering
\includegraphics*[width=75mm]{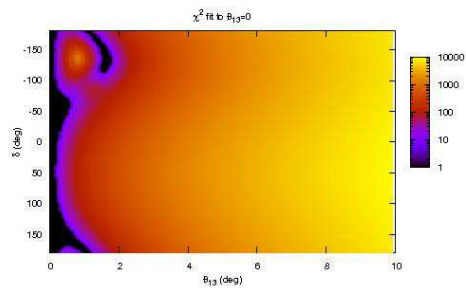}
\caption{Setup I. $\theta_{13}$ sensitivity.}
\label{sens-theta}
\end{figure}

\begin{figure}[htb]
\centering
\includegraphics*[width=75mm]{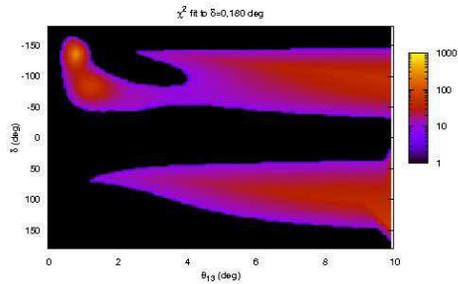}
\caption{Setup I. CPV sensitivity.}
\label{sens-delta}
\end{figure}

The corresponding exclusion plots which define the sensitivity to discover a non-vanishing mixing $\theta_{13} \ne 0$  and  CP violation $\delta \ne 0$ are presented in Fig.~\ref{sens-theta} and  Fig.~\ref{sens-delta} for varying confidence levels. For $99\%$~CL the sensitivity to a non-vanishing mixing is tipically around 1 degree. The corresponding sensitivity ($99\%$~CL) to see CP violation becomes significant for $\theta_{13}>4^o$ with values of  the phase $\delta$ around $30^o$ to be distinguished from zero.

\begin{figure}[htb]
\centering
\includegraphics*[width=55mm,angle=-90]{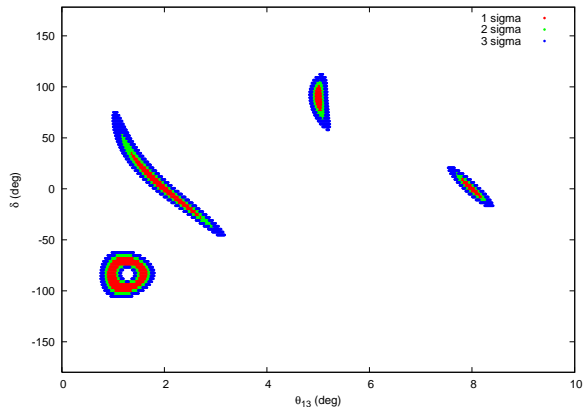}
\caption{Setup II. Fit for $(\theta_{13}, \delta)$ from statistical distribution.}
\label{fit-EC}
\end{figure}

In the case of Setup II the longer baseline for $\gamma=195$ leads to a value of $E/L$ well inside the second oscillation. In that case the associated strip in the ($\theta_{13}$, $\delta$) plane has a more pronounced curvature, so that the two parameters can be better disantangled. The statistical distribution generated for some assumed values of ($\theta_{13}$, $\delta$) has been fitted and the $\chi^2$ values obtained. The results are given in Fig.~\ref{fit-EC}. Qualitatively, one notices that the precision in the mixing is somewhat (but no much) better than that in Setup I. On the contrary, the precision reachable for the CP phase is much better than that for Setup I. One should emphasize that this improvement in the CP phase has been obtained with the neutrino channel only, using two appropriate different energies. 

\begin{figure}[htb]
\centering
\includegraphics*[width=60mm,angle=-90]{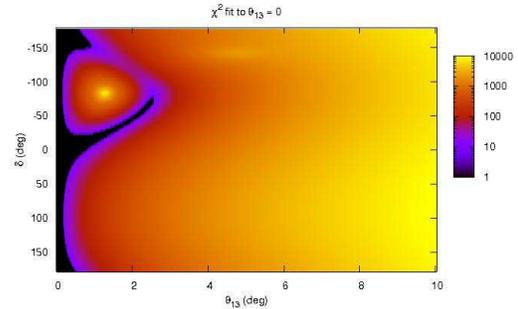}
\caption{Setup II.  $\theta_{13}$ sensitivity.}
\label{sen-EC-theta}
\end{figure}

\begin{figure}[htb]
\centering
\includegraphics*[width=60mm, angle=-90]{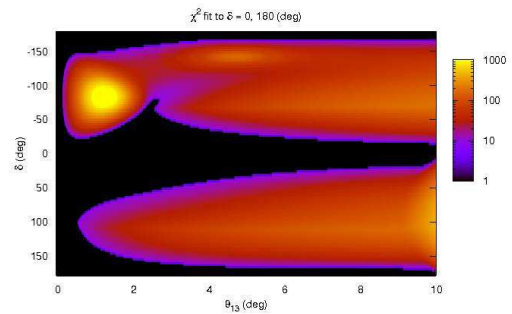}
\caption{Setup II. CPV sensitivity for the statistical distribution depending on two parameters ($\theta_{13}$ and $\delta$).}
\label{sen-EC-delta}
\end{figure}

The corresponding exclusion plots which define the sensitivities to discover a non-vanishing mixing $\theta_{13} \ne 0$ and  CP violation $\delta \ne 0$ are presented in Fig.~\ref{sen-EC-theta} and Fig.~\ref{sen-EC-delta} for varying confidence levels. For $99\%$~CL the sensitivity to a non-vanishing mixing is, as before, significant up to around 1 degree. The corresponding sensitivity ($99\%$~CL) to see CP violation for $\theta_{13}>4^o$  becomes now  significant with values  of  the phase $\delta$ around $20^o$ to be distinguished from zero. 

\begin{figure}[htb]
\centering
\includegraphics*[width=60mm, angle=-90]{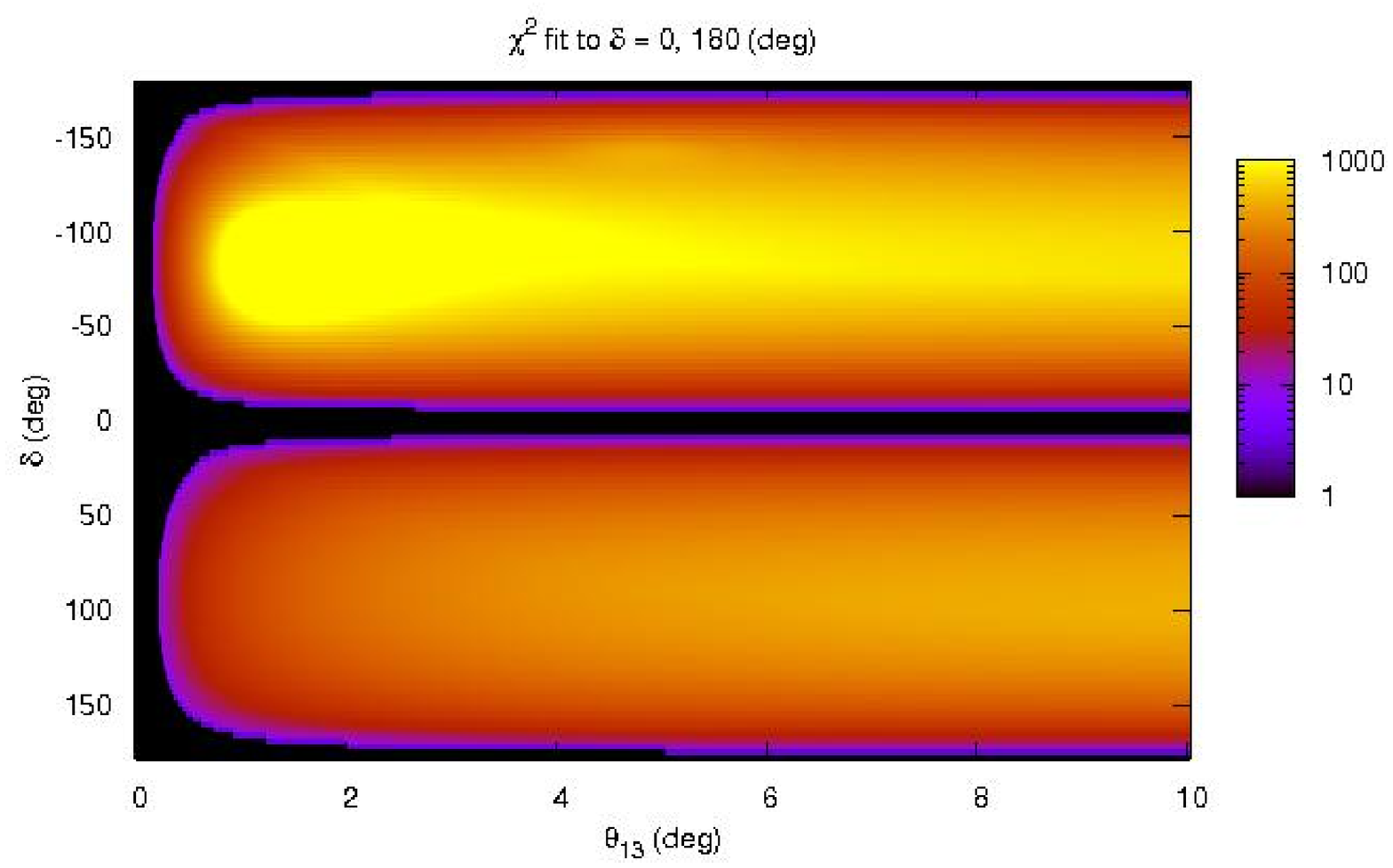}
\caption{Setup II. CPV sensitivity for the statistical distribution depending on a single parameter $\delta$, assuming previous information on $\theta_{13}$.}
\label{sen-delta-1par}
\end{figure}

At the time of the operation of this proposed Facility in Setup II it could happen that the connecting mixing $\theta_{13}$ is already known from the approved experiments for second generation neutrino oscillations, like Double CHOOZ, T2K and NOVA. To illustrate the gain obtainable in the sensitivity to discover CP violation from the previous knowledge of $\theta_{13}$ we present in Fig.~\ref{sen-delta-1par} the expected sensitivity with the  distribution of events depending on a single parameter $\delta$ for a fixed known value of $\theta_{13}$. The result is impressive: even for a mixing angle of one  degree, the CP violation sensitivity  at $99.7\%$~CL reaches values around $10^o$. 

\section{5.~Conclusions and Outlook}
\noindent The simulations of the physics output for both Beta and EC beams indicate:

1) The upgrade to higher energy ($E_p=1000$~GeV) is crucial to have a better sensitivity to CP violation, which is the main objective of the next generation neutrino oscillation experiments, iff accompanied by a longer baseline.

2) The best $E/L$ in order to have a  higher sensitivity to the mixing $U(e3)$ is not the same than that for the CP phase. Like the phase-shifts, the presence of $\delta$ is easier to observe when the energy of the neutrino beams enters  into the region of the second oscillation. The mixing is better seen around the first oscillation maximum, instead.

In particular, Setup II in EC beams, i.e., with $\gamma's$ between $195$ and $440$ and a baseline $L=650$~Km (Canfranc),  has an impressive sensitivity to CP violation, reaching  precisions around $20^o$, for $99\%$~CL, or better (if some knowledge on the value of $\theta_{13}$ is already established).  

Besides the feasibility studies for the machine, most important for physics is the study of the optimal configuration by combining low energy with high energy neutrino beams, short baseline with long baseline and/or EC monochromatic neutrinos with $^6He$ $\beta^-$ antineutrinos. 

Among the possible systematics associated with the proposed experiments one should  define a program to determine independently the relevant cross sections of electron and muon neutrinos and antineutrinos with water in the relevant energy region from several hundreds of MeV's to 1 GeV or so.

The result of the synergy of Neutrino Physics with Nuclear Physics (EURISOL) and LHC Physics (SPS upgrade) for the Facility at CERN could be completed with the synergy with Astroparticle Physics for the Detector, which could be  common to neutrino oscillation studies with terrestrial beams, atmospheric neutrinos (sensitive to the neutrino mass hierarchy through matter effects \cite{Bernabeu:1999ct}), Supernova neutrinos and Proton decay. 

The analysis shown in this paper shows that the proposals discussed here merit $R \& D$ studies in the immediate future for all their ingredients: Facility, Detector and Physics.

\end{document}